\documentclass[prl,twocolumn,preprintnumbers,amsmath,amssymb,a4paper,superscriptaddress,floatfix]{revtex4}

\usepackage{graphicx}
\DeclareGraphicsExtensions{.eps}

\pdfoutput=0

\begin{document}

\title{The Solvophobic Solvation and Interaction of Small Apolar Particles
  in Imidazolium-Based Ionic Liquids is Characterized by
  Enthalpy-/Entropy-Compensation }

\author{Dietmar Paschek} 
\email{dietmar.paschek@tu-dortmund.de}
\affiliation{Physikalische Chemie, Fakult\"at Chemie, TU Dortmund, Otto-Hahn-Str. 6, 
  D-44221 Dortmund, Germany}
\author{Thorsten K\"oddermann}
\affiliation{Physikalische und Theoretische Chemie, Institut f\"ur Chemie, 
  Universit\"at Rostock, Dr.-Lorenz-Weg 1, D-18059 Rostock, Germany}
\author{Ralf Ludwig}
\affiliation{Physikalische und Theoretische Chemie, Institut f\"ur Chemie, 
  Universit\"at Rostock, Dr.-Lorenz-Weg 1, D-18059 Rostock, Germany}
\affiliation{Leibniz Institut f\"ur Katalyse an der Universit\"at Rostock,
Albert-Einstein-Str. 29a, D-18059 Rostock, Germany}

\date{\today}

\begin{abstract}
We report results of molecular dynamics simulations characterizing the
solvation and interaction of small apolar particles such as methane and
Xenon in imidazolium-based ionic liquids (ILs). 
The simulations are able to reproduce semi-quantitatively 
the anomalous temperature dependence of the  solubility 
of apolar particles in the infinite dilution regime.
We observe that the ``solvophobic solvation'' of small apolar
particles in ILs is governed by compensating entropic and
enthalpic contributions, very much like the 
hydrophobic hydration of small apolar particles in liquid water.
In addition, our simulations clearly indicate
that the solvent mediated interaction of apolar particles dissolved
in ILs is similarly
driven by compensating enthalpic/entropic contributions,
making the ``solvophobic interaction''
thermodynamically analogous to the hydrophobic interaction.
\end{abstract}
\maketitle

Ionic liquids (ILs)  are a new class of solvents for use
in environmentally benign industrial processes and are seen
as alternative to toxic volatile organic compounds 
\cite{WasserscheidWelton,Rogers:2003,Endres:2006}.
The ionic nature of ILs has important consequences for the structure
of the liquid 
on the nanoscopic level \cite{Lopes:2006,Padua:2007}.  
Spectroscopic evidence is suggesting the presence and importance of the formation
of intermolecular cation/anion hydrogen bonds \cite{Koeddermann:2006}.
Favorable and specific  ion/cation interactions seem
to induce the formation of a persistent anion/cation network,
 which has been
shown to be quite tolerable to adding both polar and apolar particles
\cite{Rebelo:2007}. 
Quite recently, systematic measurements of the infinite dilution properties
for a number of gases, including methane, carbon dioxide, as well
as the noble gases have been reported \cite{Anthony:2002,Kamps:2003,Anthony:2005,Kumelan:2007}. 
The experimental data indicate
that imidazolium-based ILs of type 1-alkyl-3-methyl-imidazolium
bis(trifluoromethylsulfonyl)imide
(denoted as $\rm[C_nmim][NTf_2]$)
exhibit an anomalous 
temperature dependence
of the solubility of apolar compounds showing a decreasing solubility with
increasing temperature \cite{Kumelan:2007}. Moreover, it was observed that the anomalous 
behavior is found to be even
strengthened with increasing particle size.
We have recently developed an improved (nonpolarizable) all-atom
forcefield for 
imidazolium based ILs of the type $\rm[C_nmim][NTf_2]$
and have shown that a number of thermodynamical and dynamical properties
of the pure IL could be reproduced almost quantitatively
\cite{Koeddermann:2007}.
Here we show that our forcefield is also capable of semi-quantitatively
describing the solvation behavior of small apolar particles.
More importantly, we clearly demonstrate
that the solvation is characterized 
by an enthalpy-/entropy-compensation-effect, qualitatively similar to
the behavior of the hydrophobic hydration of small apolar particles
in liquid water \cite{PrattRev:2002,Southall:2002,Widom:2003,Chandler:2005}.
In addition, we determine for the first time the existence
of a ``solvophobic interaction'' of apolar particles in ILs 
which is stabilized by entropic and counter-balanced
by enthalpic contributions, similar to the hydrophobic interaction
of small solutes in water.
\begin{figure}
  \centering
  \footnotesize
  \centering
  \includegraphics[angle=0,width=4.5cm]{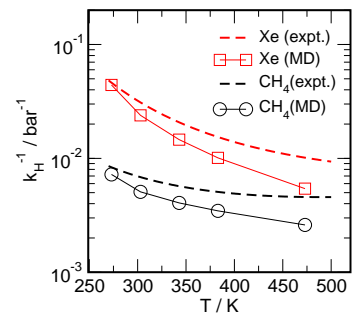}
  \caption{
    Solubility (here given as inverse Henry's constant
    for the case of infinite dilution
    with $k_H^{-1}\!=\! \exp\left[ \beta\,\mu_{ex,\rm Gas}^l\right]/
    \rho^l_{\rm IL} R T $ \cite{Kennan:90})
    of Methane and Xenon in  $\rm[C_6mim][NTf_2]$ 
    at atmospheric pressure conditions.
    The symbols indicate data obtained from MD simulations using
    our IL-forcefield \cite{Koeddermann:2007} determined from
    the potential distribution theorem \cite{Widom:63}. 
    The experimental data is according to Maurer et al. \cite{Kumelan:2007}.
    }
  \label{fig:01}
\end{figure}
\begin{figure*}
  \centering
  \footnotesize
  \centering
  \includegraphics[angle=0,width=14.0cm]{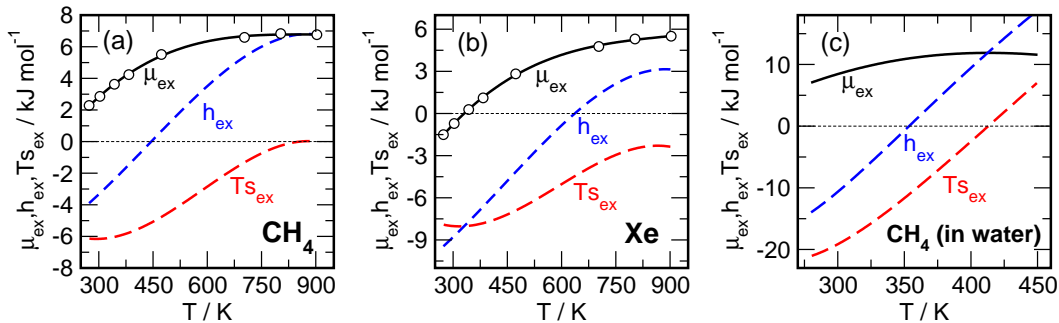}
  \caption{
    Simulated excess chemical potential $\mu_{ex}$, 
    as well as its enthalpic and entropic
    contributions $\mu_{ex}\!=\!h_{ex}-Ts_{ex}$
    of (a) Methane and (b) Xenon in $\rm[C_2mim][NTf_2]$, 
    as well as (c) Methane in water (all at atmospheric
    pressure conditions). The symbols represent the
    data obtained from the simulations applying the potential distribution
    theorem \cite{Widom:63}. 
    Enthalpic and entropic
    contributions in a,b were derived
    from fitting the $\mu_{ex}(T)$-data to a third order polynome (shown as black
    solid line \cite{MUFIT}). The experimental data for the solvation of Methane in water 
    shown in (c) are according to Refs. \cite{Prini:89,Wagner:2002}.}
  \label{fig:02}
\end{figure*}
\begin{figure}
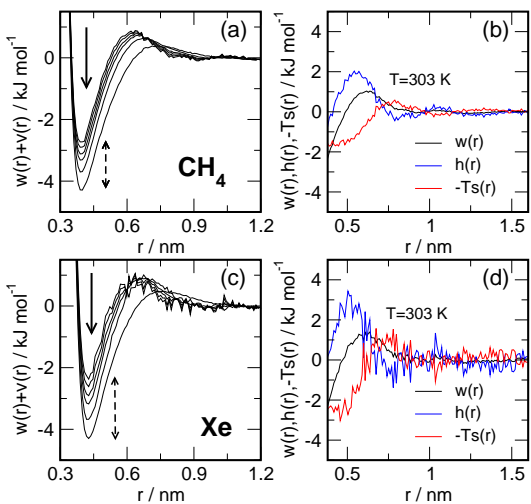

  \centering
  \footnotesize
  \centering
  \includegraphics[angle=0,width=7.0cm]{FIG03ab}
  \includegraphics[angle=0,width=7.0cm]{FIG03cd}
  \caption{
    a,c) profile of free energy $\Delta\mu_{ex}(r)=w(r)+v(r)$ for 
    the association of two Methane (a) and
    Xenon (c) particles with increasing temperature. 
    Here $v(r)$ is the intermolecular potential, whereas
    $w(r)$ denotes the solvent contribution to the free energy profile.
    The shown temperatures are 
    $273\,\mbox{K}$, $303\,\mbox{K}$, $343\,\mbox{K}$, $383\,\mbox{K}$ 
    , $473\,\mbox{K}$, and $703\,\mbox{K}$. The arrow indicates increasing
    temperature.
    b,d) solvent contribution to the profile of the free energy, as well as the enthalpic
    and entropic contributions obtained for $T\!=\!303\,\mbox{K}$. Methane (b). Xenon (d).}
  \label{fig:03}
\end{figure}

We perform constant pressure (NPT) MD simulations of imidazolium based ILs of the
type $\rm[C_2mim][NTf_2]$ and $\rm[C_6mim][NTf_2]$ at a pressure of $1\,\mbox{bar}$
over the temperature range of $273\,\mbox{K}$ up to $907\,\mbox{K}$ \cite{STATES}
of system-sizes of 173 ion pairs using our IL forcefield \cite{Koeddermann:2007}.
Simulations of at least $10\,\mbox{ns}$ length are employed for each
state-point. The simulation conditions \cite{techdetails} are similar as in
Ref. \cite{Koeddermann:2007}. All simulations
are performed with the Gromacs 3.2 simulation program \cite{gmxpaper}.
We study the solvation of small apolar particles for the
for the case of Methane and Xenon represented by single
Lennard-Jones spheres with parameters
($\sigma_{\mbox{Me-Me}}\!=\!0.3730\,\mbox{nm}$,
$\epsilon_{\mbox{Me-Me}}/k\!=\!147.5\,\mbox{K}$, and
$\sigma_{\mbox{Xe-Xe}}\!=\!0.3975\,\mbox{nm}$,
$\epsilon_{\mbox{Xe-Xe}}/k\!=\!214.7\,\mbox{K}$)
\cite{Paschek:2004:1}
applying Lorentz-Berthelot mixing rules \cite{Allen-Tildesley} for all cross-terms.
The solvation free energy per particle is given by
the excess chemical potential $\mu_{ex}$. We determined
$\mu_{ex}$ for the case of infinite dilution
a posteriori from the IL MD-trajectories applying
Widom's potential distribution theorem \cite{Widom:63} with
$\mu_{ex}\;= \; - kT \ln \left< V \exp(-\beta\,\Phi(\vec{r})) \right>/\left<V\right>$.
Here is $\beta\!=\!1/kT$, $V$ the volume of the simulation box, 
and $\Phi(\vec{r})$ is the energy of a randomly inserted (gas) test-particle
at position $\vec{r}$. 
The brackets $\left< \ldots\right>$ indicate isobaric isothermal sampling
as well as sampling over many different positions $\vec{r}$. More details
about the calculation are available in Ref. \cite{Paschek:2004:1}.
The solubility of the gas-particles 
is given as the inverse Henry's constant 
$k_H^{-1}\!=\! \exp\left[ \beta\,\mu_{ex,\rm Gas}^l\right]/
    \rho^l_{\rm IL} R T $ \cite{Kennan:90})
where $,\mu_{ex,\rm Gas}^l$ is the chemical potential of the solute
in the liquid phase and $\rho^l_{\rm IL}$ is the average number density
of the ionic liquid. Note that the number density of
neutral ion-pairs is used here.

Figure 1 compares the solubility
of Methane and Xenon in $\rm[C_6mim][NTf_2]$ as obtained from our MD simulations with
solubility data recently published by Maurer et al. \cite{Kumelan:2007}.
We would like to emphasize that without any refinement of the potential
parameters, the solubility-data is quite satisfactorily reproduced.
For low temperatures the agreement is almost quantitatively.
In addition, the particle-size dependence is well described. We would
particularly like to stress the  fact that both simulation and experiment 
indicate an {\em anomalous} temperature dependence, 
showing a decreasing solubility with increasing temperature,
very much resembling the solvation of apolar gases in water \cite{Wilhelm:77}.
It is the purpose of this letter to demonstrate
that the analogy to the ``hydrophobic effect'' is much 
more deep-rooted thermodynamically.

In Figure 2 we show the excess chemical potentials of Methane and Xenon
dissolved in $\rm [C_2mim][NTf_2]$. For comparison the excess chemical
potential of Methane in liquid water is given.
The increasing chemical potential
with increasing temperature equivalent to the anomalous solvation.
In analogy to the procedure  used in Ref. \cite{Paschek:2004:1}
the data points shown in Figure 2 were fitted to a third order
polynome  \cite{MUFIT} and the corresponding entropic and
enthalpic contributions were determined from the 
temperature dependence of the fitted $\mu_{ex}(T)$
with $s_{ex}\!=\!-(\partial \mu_{ex}/\partial T)_{T,P}$
and $h_{ex}=\mu_{ex}+Ts_{ex}$. From Figure 2a,b it is evident
that for the low temperature regime the negative heat of solvation $h_{ex}$ is
counter compensated by a negative solvation entropy $s_{ex}$. Moreover,
the heat of solvation exhibits a positive slope, revealing
a positive solvation heat capacity contribution of about 
$22\,\mbox{J}\mbox{K}^{-1}\mbox{mol}^{-1}$  for Methane and about
$26\,\mbox{J}\mbox{K}^{-1}\mbox{mol}^{-1}$  for the temperature interval
between $300\,\mbox{K}$ and $400\,\mbox{K}$. Both, the entropy/enthalpy
compensation effect, as well as the positive heat capacity are also qualitative
signatures of the hydrophobic hydration of small apolar particles in
water \cite{Silverstein:2000,Southall:2002}. However, 
the solvation heat capacity is about a factor
of five to six smaller, and the corresponding solvation entropies are
about a factor of three to four smaller compared to the solvation in
water (compare with Figure 2c and data in Ref. \cite{Paschek:2004:1}). 
In addition,
the maximum of  $\mu_{ex}(T)$, which is observed in water around
$410\,\mbox{K}$ to $420\,\mbox{K}$ is shifted to about
$900\,\mbox{K}$ to $1000\,\mbox{K}$ for the case of $\rm[C_2mim][NTf_2]$.
The latter values are of course hypothetical, since the 
real ILs are not chemically stable under those extreme conditions.
Our simulations indicate the anomalous
solvation behavior of ILs is simply stretched out on a much broader 
temperature scale compared to  water.

In addition to the solvation behavior, 
we also determine the solvent mediated interaction between two (identical) gas
particles. Therefore we calculate the profile of free energy for the association process by
applying the potential distribution theorem \cite{Widom:63}
with $w(r) \!=\! -kT \ln \left< V \exp(-\beta\,\Phi(\vec{r}_1,\vec{r}_2)) \delta
  (|\vec{r}_1 - \vec{r}_2|-r) \right>/\left<V\right> -
2 \mu_{ex}$. Here $\Phi(\vec{r}_1,\vec{r}_2)$ is the energy of randomly
inserting two gas particles. $\mu_{ex}$ is the excess chemical potential
of the individual gas particles. 
The solvent mediated interaction $w(r)$ is related to the gas-gas pair distribution
function $g(r)$ according to $ -kT \ln g(r)=w(r)+v(r)$, where $v(r)$ is the
intermolecular pair potential between the two gas particles.
$w(r)$ is also sometimes referred to a ``cavity potential''
\cite{Ben-Naim:Hydrophobic}. Similar to the excess chemical potential, $w(r)$
was calculated a posteriori from stored trajectory data using a Monte
Carlo procedure.
The calculated profiles of free energy, as well as the 
corresponding enthalpic and entropic contributions
for the association of two Methane and Xenon particles
are shown in Figure 3. The minimum of the profile of free energy 
shown in Figure 3a,c represents
the state where two particles are in close contact. 
Differing from the the profile of free energy for small hydrophobic particles
dissolved in water, there is no pronounced second minimum existing here. 
Hence, the presence of a clearly defined solvent separated
state is missing, which is likely to be related to the much
larger size of solvent molecules involved here compared to water.
However, in parallel to the behavior in water, we do observe
a deepening of the first minimum with increasing temperature.
When comparing the well depth of the  minima in Figure 3a,c, Xenon
exhibits stronger temperature effect than Methane.
To determine enthalpic and entropic contributions, we have fitted
each $r$-histogram-point of the set of $w(r,T)$-histograms to a third 
order polynome with respect to $T$. The corresponding profiles
calculated for $T\!=\!303\,\mbox{K}$ are shown in Figure 3b,d. 
The contact-state for both, Methane and Xenon
is stabilized by the entropy part, whereas the enthalpic part
mostly destabilizing. A simple explanation is based on the
assumption that solvation enthalpy/entropy is mostly due to
changes of the solvent in the first solvation shell, which
is represented by the solvent accessible
surface (SAS). In the contact state the SAS is minimized, leading
to negative net entropy and positive net enthalpy for the association
of two particles as depicted in Figure 4. Hence the association of apolar
particles in ILs is driven by the tendency to {\em minimize the
solvation entropy-penalty} (the phrase has been borrowed from
Haymet et al. \cite{Haymet:96})
similar to what has been found for water \cite{Haymet:96,Southall:2002}.
Consequently, the larger temperature dependence of the well depth of the
profile of free energy  observed for Xenon in Figure 3,a,c 
is simply a consequence of its larger solvation entropy (compare
data in Figures 2a,b).
\begin{figure}[!t]
  \centering
  \includegraphics[angle=0,width=7.0cm]{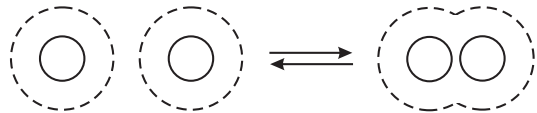}
  \caption{Schematic diagram of the ``solvophobic
    association'' process. The contact
    configuration is stabilized with increasing temperatures
    by minimizing the entropy penalty.}
  \label{fig:04}
\end{figure}

Our simulations reveal interesting new insights into the 
solvation and interaction of apolar particles in ionic liquids. However,
many questions remain unanswered.
First of all, we have not shown what is exactly causing
the entropy penalty of the ``solovophobic solvation'' 
in ILs. It might well be that the effect is similar to the hydrophobic
hydration in water, where the entropy penalty has been
largely attributed to the orientational bias put on
the water molecules while trying to keep the
hydrogen bond network around an apolar particle mostly intact \cite{Southall:2002}.
Since the formation of intermolecular hydrogen-bonds is also an important
feature in ILs \cite{Koeddermann:2006}, a similar
mechanism might also apply here. 
However, it might also well be that 
the tendency to maximize anion-cation contacts in ILs \cite{Padua:2007}
is introducing an ordering constraint in the
solvation shell of an apolar particle and thus causing the entropy penalty.
In addition, for the case of water it has been 
recognized recently that the solvation of small apolar particles,
which is ``entropy dominated'',
is  different than for large scale
particles, which is ``enthalpy dominated''. Hence there
has to be a crossover, which has been placed on the $<1\,\mbox{nm}$-scale 
as proposed by the theory of Lum et al. \cite{Lum:1999, Chandler:2005}.
The predicted size of the crossover-lengthscales were confirmed recently
by computer simulations of Rajamani et al. \cite{Rajamani:2005}
and by simulation based scaled-particle theory \cite{Ashbaugh:2006, Ashbaugh:2007}.
Given the larger size of the IL molecules and considering the importance of
maintaining anion/cation contacts, the crossover-lengthscale might
be shifted to larger values for the case of ILs.
Considering that the anomalous solvation behavior
of gases increases with solute molecule-size \cite{Anthony:2005,Kumelan:2007},
the poor solubility of proteins in most
pure ILs  \cite{Klembt_ILbook} might 
be a consequence of a pronounced solvation-entropy effect.

\noindent{\bf Acknowledgments.}
DP acknowledges gratefully support from the Deutsche Forschungsgemeinschaft
(DFG SPP 1155) and from TU Dortmund. 
RL acknowledges financing by the state of Mecklenburg-Vorpommern,  
and partial support by the Deutsche Forschungsgemeinschaft.

\end{document}